\documentstyle[11pt,newpasp,twoside,epsf]{article}
\begin{document}
\title{Ten years of nearby galaxies investigation with the
       Multi-Pupil Spectrograph of the 6m telescope}
\author{Sil'chenko Olga K.}
\affil{Sternberg Astronomical Institute, University av. 13, Moscow
119899, Russia; olga@sai.msu.su}

\begin{abstract}
We describe the history and results of our study of
early-type disk galaxies with the Multi-Pupil
Spectrograph of the 6m telescope (MPFS). With the new data we confirm
the discovery of chemically distinct nuclei reported in 1992;
now such nuclei are found in two dozens galaxies. The structure of
chemically distinct features formed probably in secondary nuclear
star-formation bursts seems to be complex and closely related to an
appearance of separate circumnuclear disks. As by-product, we have
also found inner polar gaseous disks in 4 non-interacting, regular
spiral galaxies; this phenomenon may be related to triaxiality.
Some preliminary statistical conclusions
on the stellar population properties in the nuclei and in the bulges
of S0 galaxies are given.
\end{abstract}

\section{Introduction}

The Multi-Pupil Spectrograph which works at the prime focus of the
6m telescope for 13 years was born due to close acquaintance of
Prof. Afanasiev (during his stay in Marseille
Observatory at the beginning of 1988) with the ideas of Prof.
Courtes. The first variant of MPFS (Afanasiev et al. 1990) used
fibers after the lens array to transmit micropupils to the entrance
of a classical `long-slit' spectrograph. The detector was IPCS
 $512 \times 512$. The fibers were quite imperfect, due to them
the spectral instrumental profile was ugly. We started our scientific
observations with the Multi-Pupil Fiber Spectrograph (MPFS) of the
6m telescope in 1989. Our initial preference was gas
kinematics, but we had obtained a typical velocity error
of more than 50 km/s and had understood that kinematical tasks cannot
be effectively made with that MPFS. So we looked for
another scientific goal. The same year we published a list
of spiral galaxies with kinematically decoupled nuclei that we thought
to be unresolved central mass condensations surrounded by circumnuclear
gaseous disks (Afanasiev, Sil'chenko, \&\ Zasov 1989). The idea had arisen
to study stellar populations of these decoupled nuclei and to see if they
differ from the stellar populations of the surrounding bulges.
During the end of 1989 and the first half of 1990 we observed a sample of
10 nearby disk galaxies, including 5 spiral galaxies with kinematically
decoupled centers. We measured radial variations of
azimuthally-averaged equivalent widths of the absorption lines
MgIb$\lambda$5175 and FeI$\lambda$5270 and had immediately detected
chemically distinct nuclei in 6 galaxies from 10, including
3 galaxies with kinematically decoupled centers. The
drops of EW(Mg) between the unresolved nuclei and their
outskirts being calibrated into metallicity terms implied that the
nuclei are more metal-rich by almost an order, if the ages of the
stellar populations are equal. These results were published in 1992
(Sil'chenko, Afanasiev, \&\ Vlasyuk 1992). Later the spectrograph
was modified more than once: in 1992 the fibers were removed, and MPFS
began to work in a classic `TIGER-like' mode (Bacon et al. 1995),
in 1994 the first CCD had come into operation. In fact, MPFS
was three different spectrographs during its whole life: the
Multi-Pupil Fiber Spectrograph (1989--1992), the Multi-Pupil
Field Spectrograph (1993--1998, Afanasiev et al. 1996), and again
the Multi-Pupil Fiber Spectrograph (1998--now,
http://www.sao.ru/$\sim$gafan/devices/mpfs/).
The present version of MPFS is equipped with CCD TK $1024 \times 1024$;
it allows to expose at once 240 spectra from an area of
$16\arcsec \times 15\arcsec$. Besides a target, it accumulates
16 spectra of the blank sky area on the same frame so sky subtraction
is now straightforward. The main reduction procedures are written
by Prof. Afanasiev under IDL. The data quality is now sufficient
to study 2D distributions of various characteristics.
From a single MPFS exposure we derive: continuum brightness and
emission-line intensity maps, stellar line-of-sight velocity field,
stellar velocity dispersion field, line-of-sight velocity field
of the ionized gas, and the key issue -- maps  of Lick
absorption-line indices. More than fifty early-type disk galaxies
are already observed with the MPFS at the 6m telescope, mainly in
the frame of my program of searching for chemically distinct galactic
nuclei. More than two dozens of chemically distinct nuclei are detected.
We have also re-investigated the galaxies whose chemically distinct nuclei
were claimed in our first paper; all of them
are confirmed. Figure~1 shows our modern data on magnesium-index radial
variations in two galaxies with chemically distinct nuclei, NGC~1023 and
NGC~7332, and in one galaxy without such nucleus, NGC~4550. Also,
the data obtained with the SAURON of the WHT and taken from the ING Archive
(we have reduced them with the software developped in the SAO for the second
version of the MPFS by Vlasyuk 1993) and the long-slit data of Fisher,
Franx, \&\ Illingworth (1996) are plotted for comparison.
One can see the sharp drops of the magnesium-line strength beyond the
nuclei in NGC~1023 and NGC~7332; the larger field of view provided by
the SAURON allows to trace more surely the flattening of the Mgb profiles
in the bulges, especially in NGC~1023 where magnesium gradient in the
bulge is absent. If we compare the nuclear measurements with
the bulge data extrapolation toward $R=0\arcsec$, we obtain a metallicity
difference of 2--3 times between the nucleus and the bulge under the
assumption of equal ages. So far the existence of the chemically
distinct nuclei in nearby regular disk galaxies is now firmly established.
   
\begin{figure}
\plotfiddle{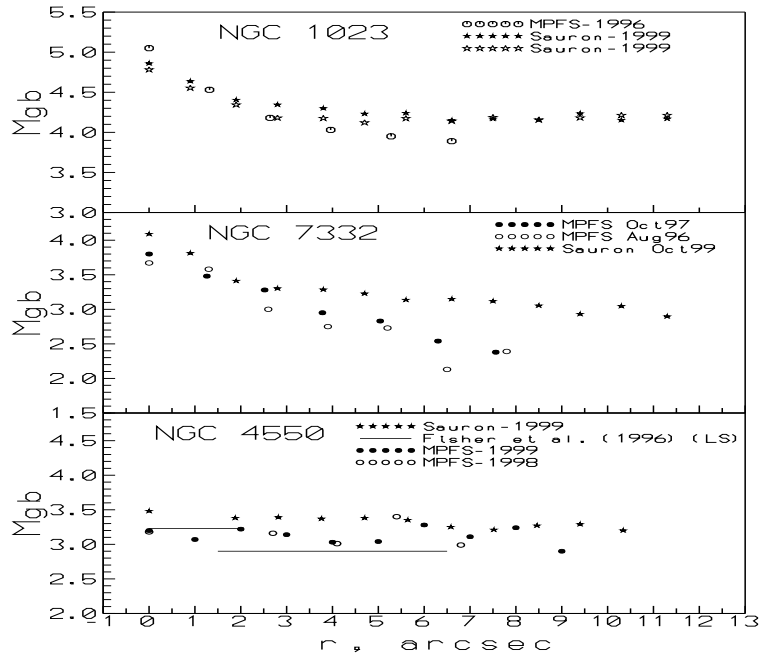}{8cm}{0}{80}{50}{-230}{-90}
\caption{Radial profiles of azimuthally-averaged Mgb index for
three galaxies according to our MPFS data, to the data from the
SAURON/WHT and to the long-slit data of Fisher et al. (1996).}
\end{figure}

\section{Chemically distinct nuclei $\equiv$ compact stellar disks?}

What is a structure -- and hence the origin -- of chemically distinct
nuclei in disk galaxies? When Bender \&\ Surma (1992)
found a spatially resolved chemically distinct core in the elliptical
NGC~4365, they decided that it is a circumnuclear
stellar disk within the giant spheroid -- mainly because
of its fast rotation. Though this hypothesis is not fully confirmed
for the other ellipticals with decoupled cores, we feel that
something like that is applicable to the chemically distinct nuclei
in disk galaxies. The latters are at the
limit of our resolution, with the radii less than $3\arcsec$,
but high-resolution HST photometric data often reveal
a presence of distinct circumnuclear disks within the bulges in the
same galaxies where we see chemically distinct nuclei. Sometimes
we can even guess a fine structure of the 2D distributions of stellar
population properties. Let us look in  detail
at two galaxies: NGC~1023 (Silchenko 1999a)
and NGC~7331 (Sil'chenko 1999b). In both galaxies the nuclei
demonstrate strong magnesium overabundance, [Mg/Fe]$\approx +0.3$. The
Mgb-index is peaked in the center and decreases radially
quasi-axisymmetrically. The maximum of the iron indices, though
located also around the nuclei, is more extended (`flat') and
strongly elongated. Isolines of the H$\beta$-index that is an age
indicator repeat the shape of the $<\mbox{Fe}>$ isolines. We should call
these quasi-resolved structures `Fe-rich disks'. The stellar
populations of the `Fe-rich disks' have a solar Mg/Fe ratio and are
younger than the nuclear ones by 2--3~Gyr: 5~Gyr vs 7~Gyr
in NGC~1023 and 5~Gyr vs 2~Gyr in NGC~7331. The bulges in both galaxies
are old, $T_{bul} \geq 15$ Gyr, and mildly magnesium-overabundant. To
explain this combination of properties, we propose a following scenario.
Several Gyrs ago a secondary star-formation burst starts over
an extended central area of a galaxy. In the very center it proceeds
more effectively (due to higher gas density?) and finishes in less than
1~Gyr -- to provide a strong magnesium overabundance of the nuclear stellar
populations. But at the edges of the circumnuclear (initially gaseous)
disks star formation continues over a few billion years resulting in a
solar Mg/Fe ratio and a younger mean stellar age. Since the whole process
is rather long, we should be able to catch some future chemically distinct
`cores'  during the non-central star formation. This scenario
explain a phenomenon of blue or even H$\alpha$-emission rings in the
early-type axisymmetric (non-barred) galaxies -- we
have found such rings in NGC~759 (Vlasyuk \&\ Sil'chenko 2000) and
NGC~80 (Sil'chenko et al., in press).

\section{Inner polar gaseous disks in regular spiral galaxies}

\begin{figure}
\plotfiddle{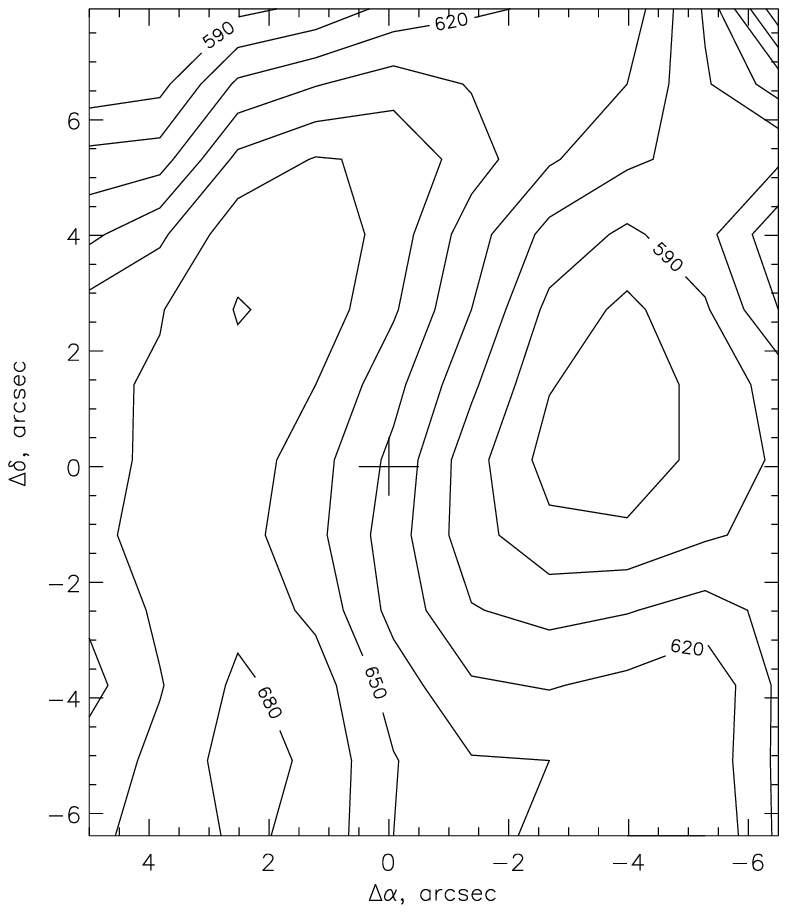}{4cm}{0}{70}{70}{-220}{-130}
\plotfiddle{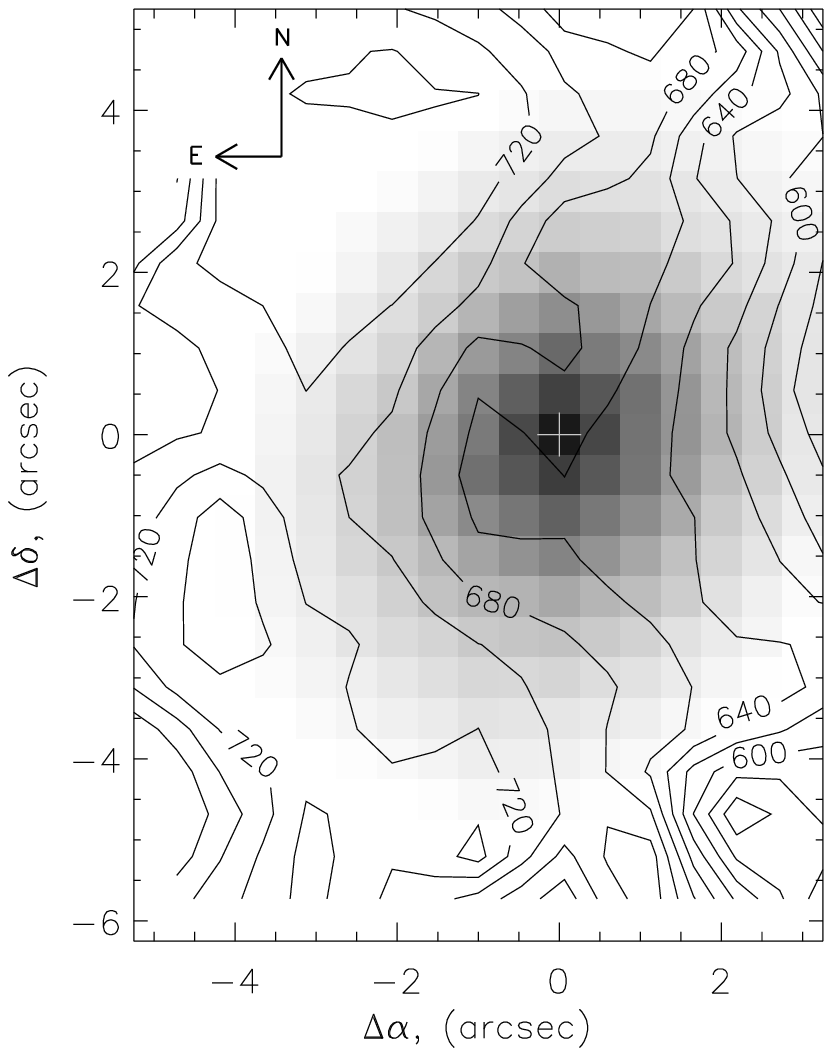}{4cm}{0}{60}{60}{-30}{35}
\plotfiddle{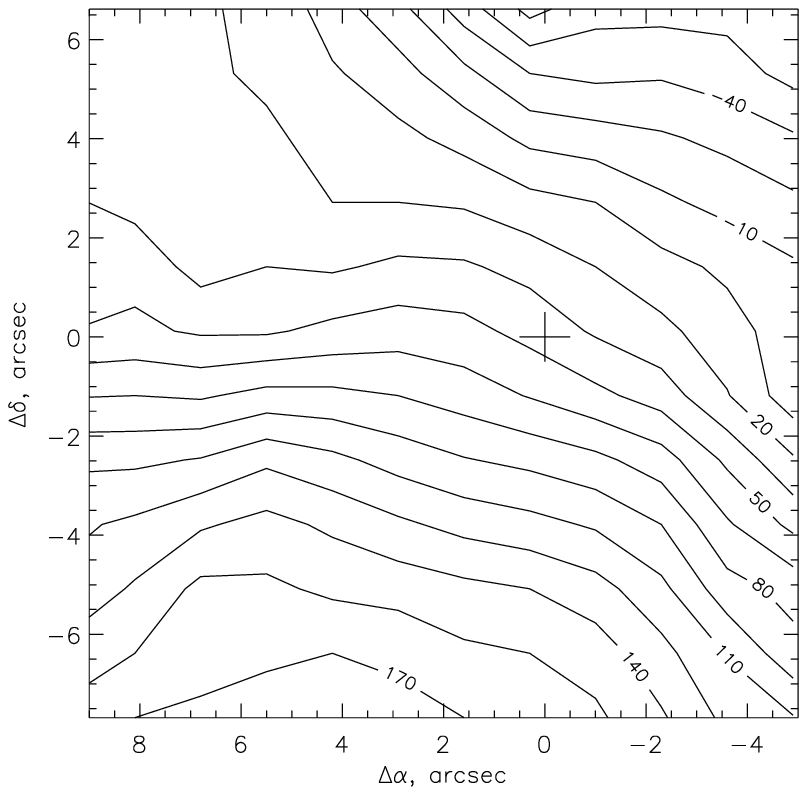}{4cm}{0}{70}{70}{-220}{-90}
\plotfiddle{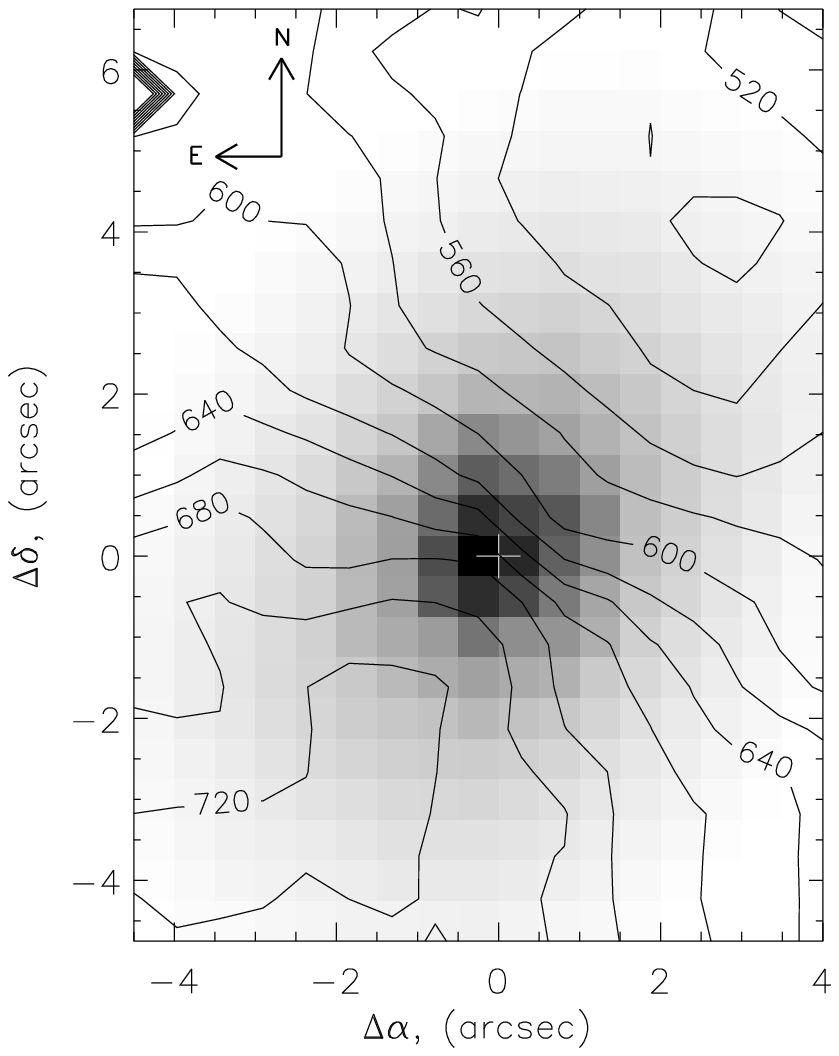}{4cm}{0}{60}{60}{-30}{75}
\caption{Isolines show line-of-sight velocity fields for the
ionized gas (top) and stars (bottom) in the center of NGC~2841
according to the MPFS data (left) and to the 2D-FIS/WHT data (right).
The crosses mark the photometric center of the galaxy.}
\end{figure}

One of our by-products during the investigation of the stellar
populations in early-type disk galaxies was discovery
of inner polar gaseous disks in non-interacting
regular spiral galaxies. There is a rare class of peculiar
galaxies with large-scale polar rings -- the example
is NGC~2685; these galaxies are supposed to accrete external gas from
another galaxy with orthogonal rotational momentum. Also some S0's
are known to have decoupled stellar and gaseous rotation
axes in their centers (Bertola, Buson, \&\ Zeilinger 1992; Bertola
et al. 1995); an origin of this decoupling is usually related to
external gas accretion as well. Meanwhile we have detected inner polar
gaseous disks, with radii of a few hundred parsecs, in spiral galaxies
possessing global HI-disks with a quite normal rotation --
these galaxies are NGC~2841, 4548, 6340, and 7217. For one of them,
NGC~2841, we have had possibility to check our MPFS results by
comparing them to the data of the 2D-FIS spectrograph operating at
the WHT. Figure~2 shows this comparison: the left plots are the
MPFS results (Sil'chenko, Vlasyuk, \&\ Burenkov 1997), the right plots
are the 2D-FIS ones. One can see that
two data sets are consistent: the gas is really rotating
directly to the stars (and in the plane orthogonal to the main
symmetry plane of the galaxy, $PA_0=150\deg$),
and even the shift of the gas rotation center by some $2\arcsec$ to
the north from the photometric center is confirmed. As the more
outer gas in all four cases belongs to the main planes
and rotates normally in these galaxies, an external origin of
the inner gas with the orthogonal momenta seems to be doubtful.
There must exist any intrinsic mechanism to put the gas onto the
polar orbit when approaching the center. Sofue \&\ Wakamatsu (1994)
give qualitative arguments that a triaxial potential, or a bar,
may easily do so. Among our galaxies with inner polar disks, only
NGC~4548 is a bona-fide barred galaxy, the others are SA. But when we
have studied photometric structure of the galaxies
in detail, we have found triaxial bulges in NGC~2841 (Afanasiev \&\
Sil'chenko 1999) and NGC~6340 (Sil'chenko 2000), and an oval inner
disk in NGC~7217 (Sil'chenko \&\ Afanasiev 2000). So
perhaps the phenomenon of the inner polar gas is indeed related
to the triaxial potential rather than to hypothetical accretion.

\section{Preliminary statistics of the stellar population properties
in the nuclei and bulges of lenticular galaxies}

\begin{figure}
\plottwo{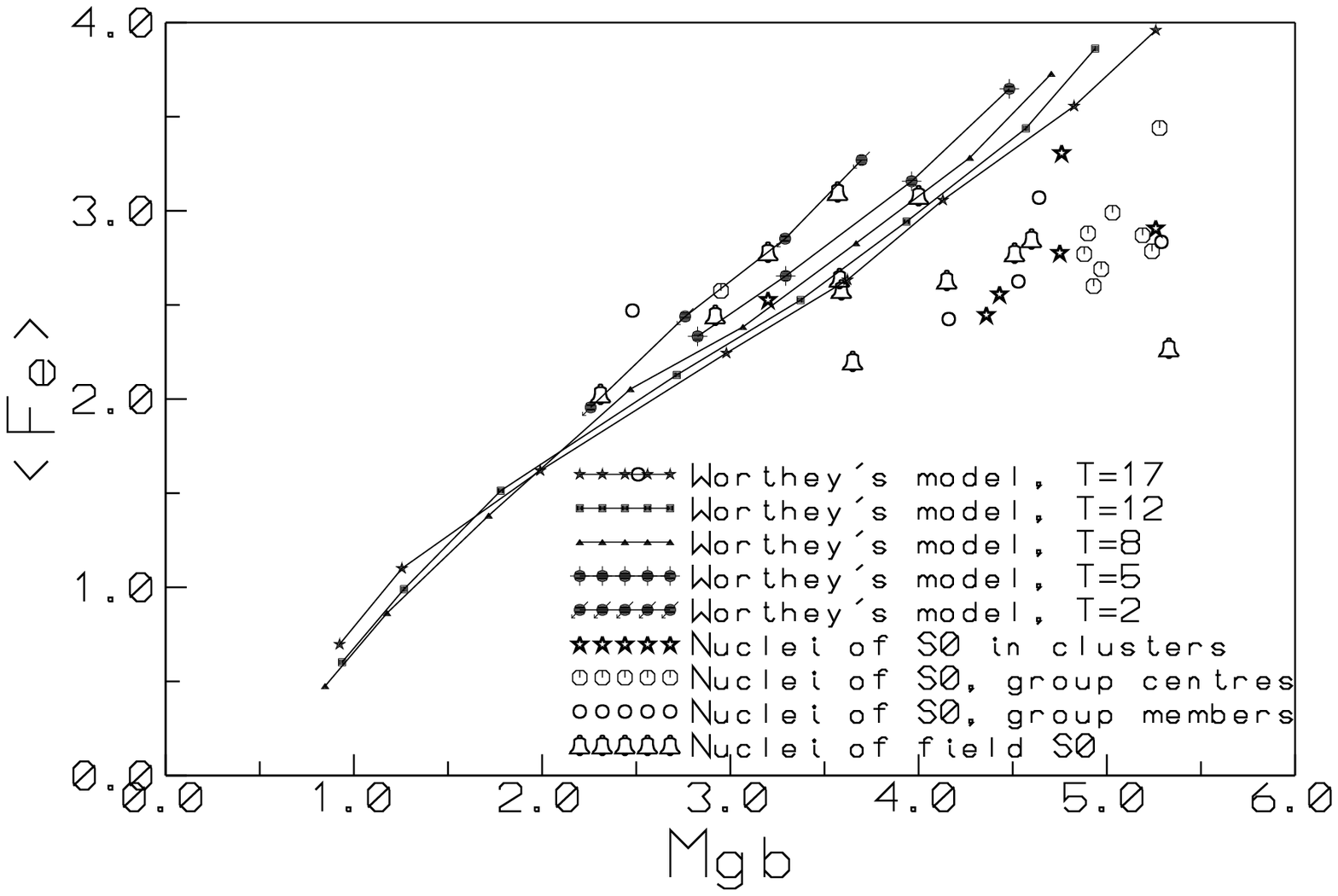}{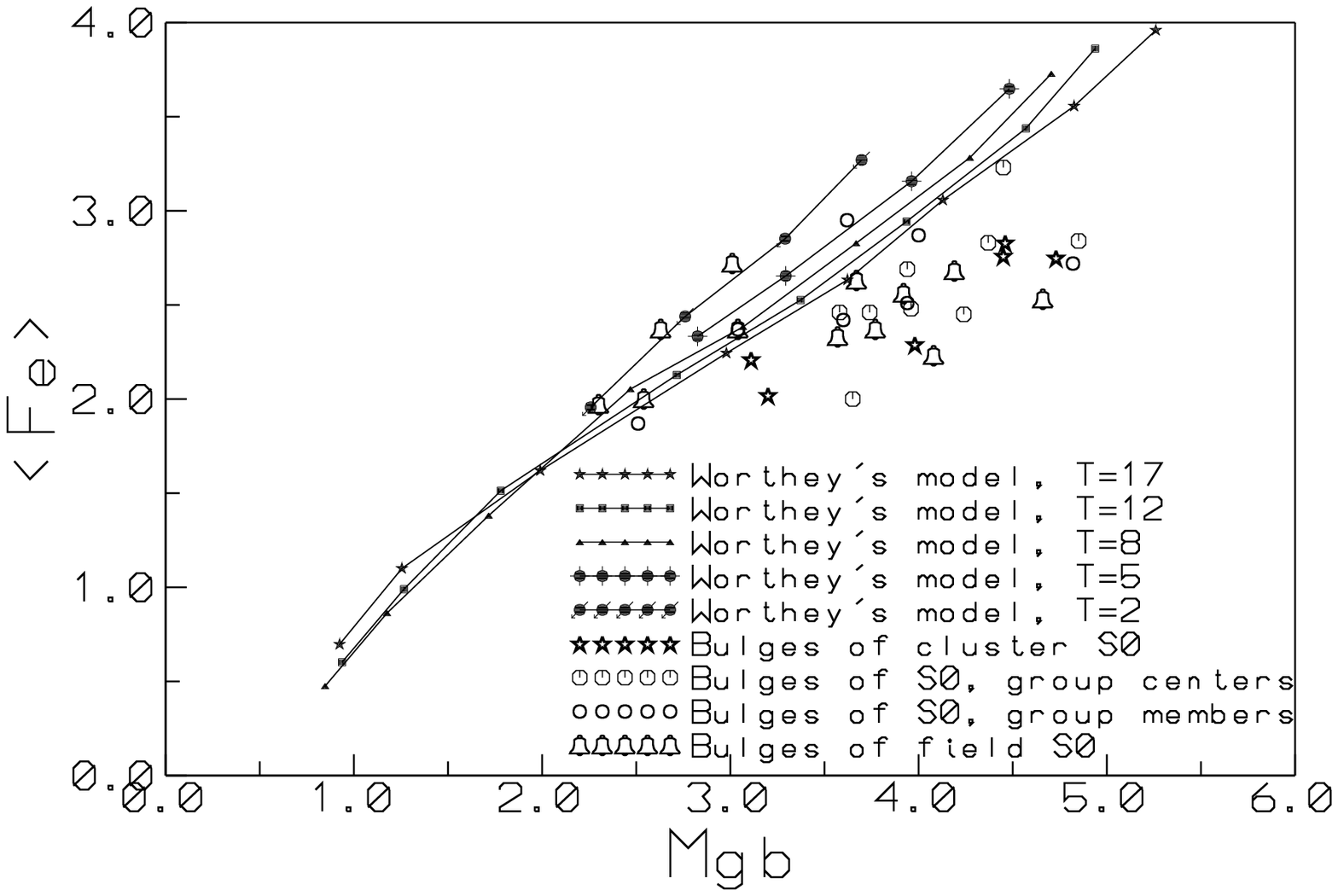}
\caption{The `magnesium versus iron' diagrams for the nuclei (left)
and the bulges (right) of the nearby lenticular galaxies. Different
large signs show the galaxies in various environment types. The small
signs connected by thin lines present stellar population models of equal
ages with [Mg/Fe]=0 from Worthey (1994); the signs marks the metallicities
-- if one takes them from the right to the left,
 the metallicities for the models are +0.50, +0.25, 0.00,
--0.22, --0.50, --1.00,--1.50, --2.00.}
\end{figure}

\begin{figure}
\plottwo{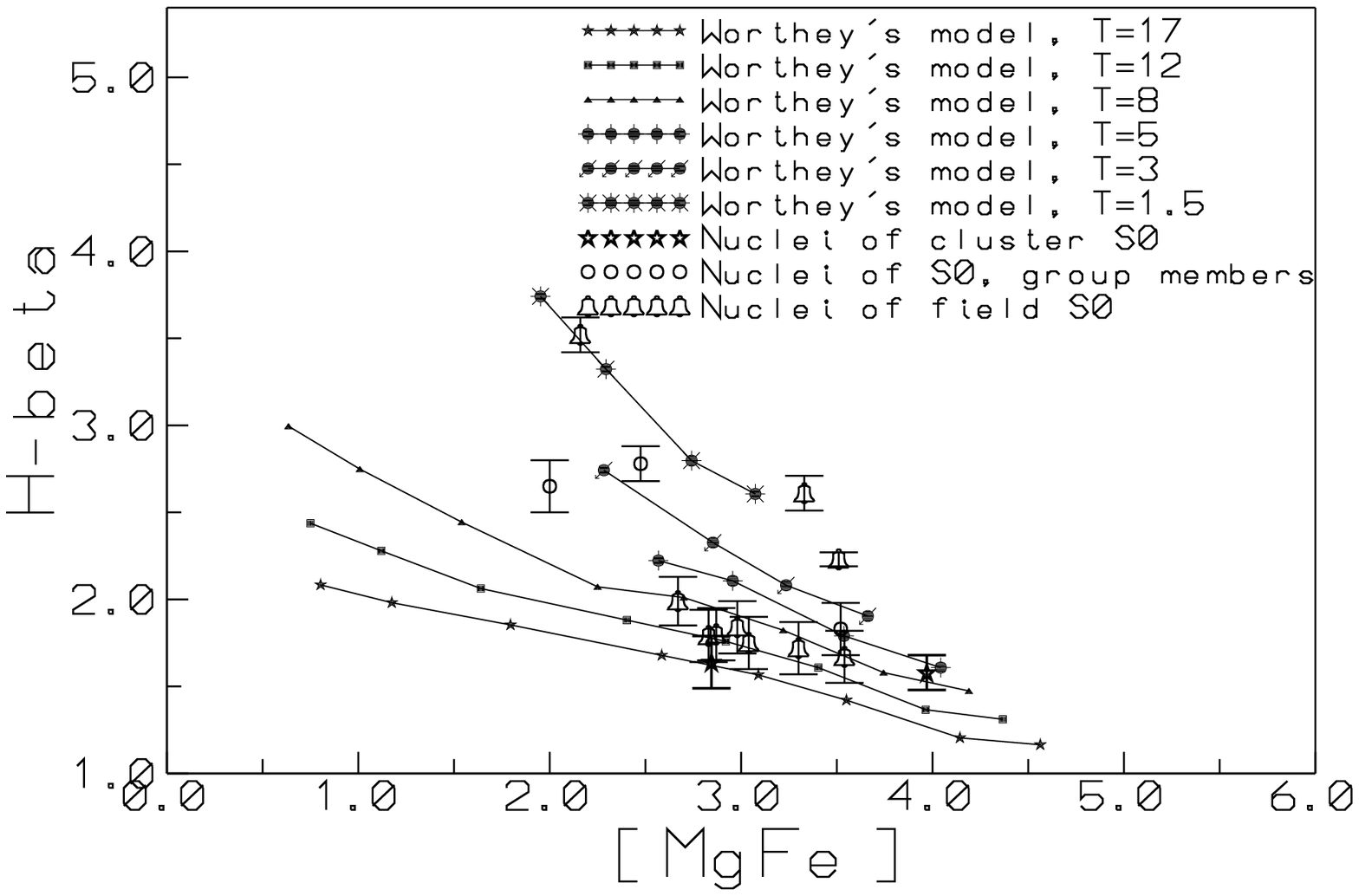}{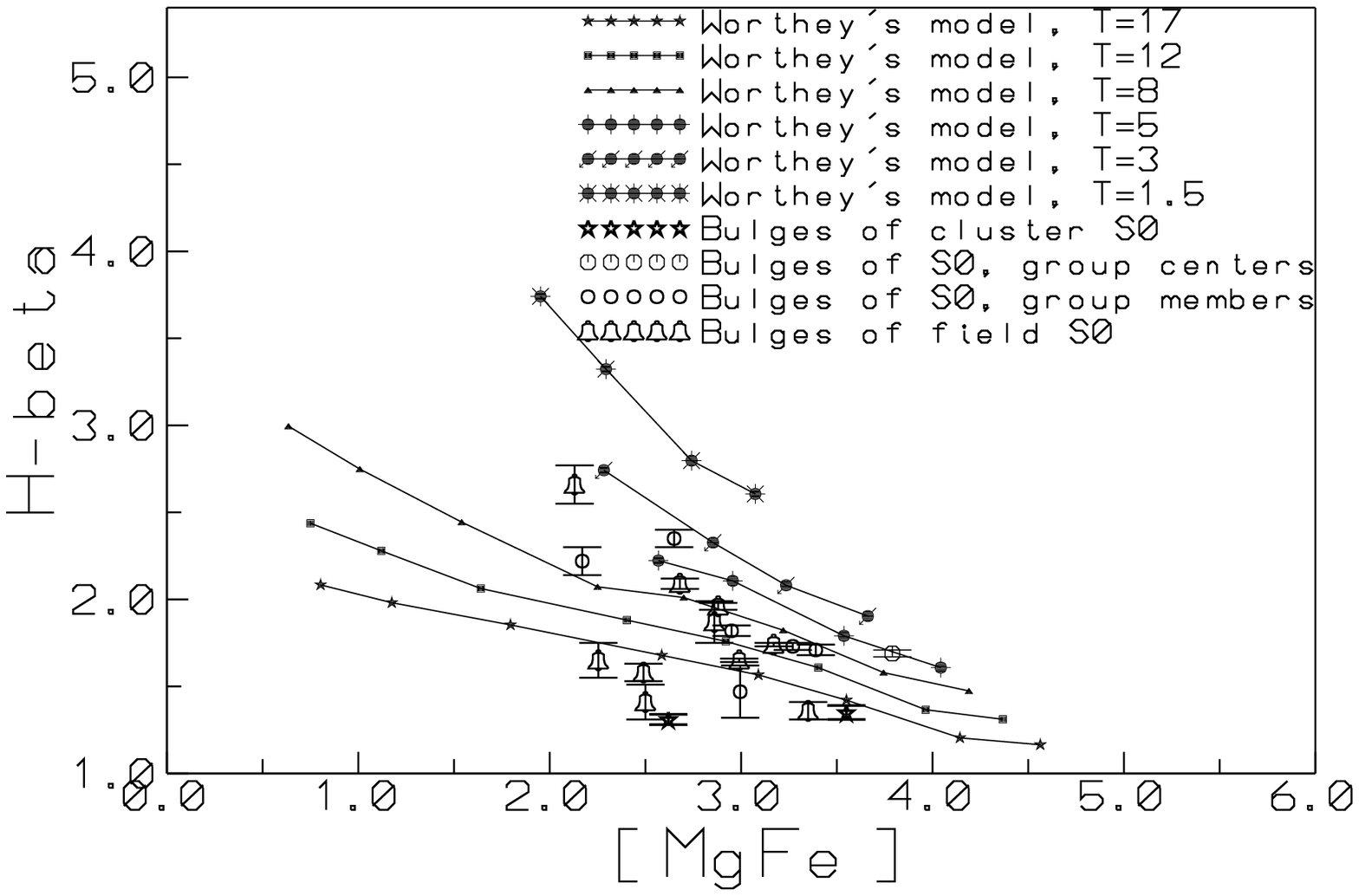}
\plottwo{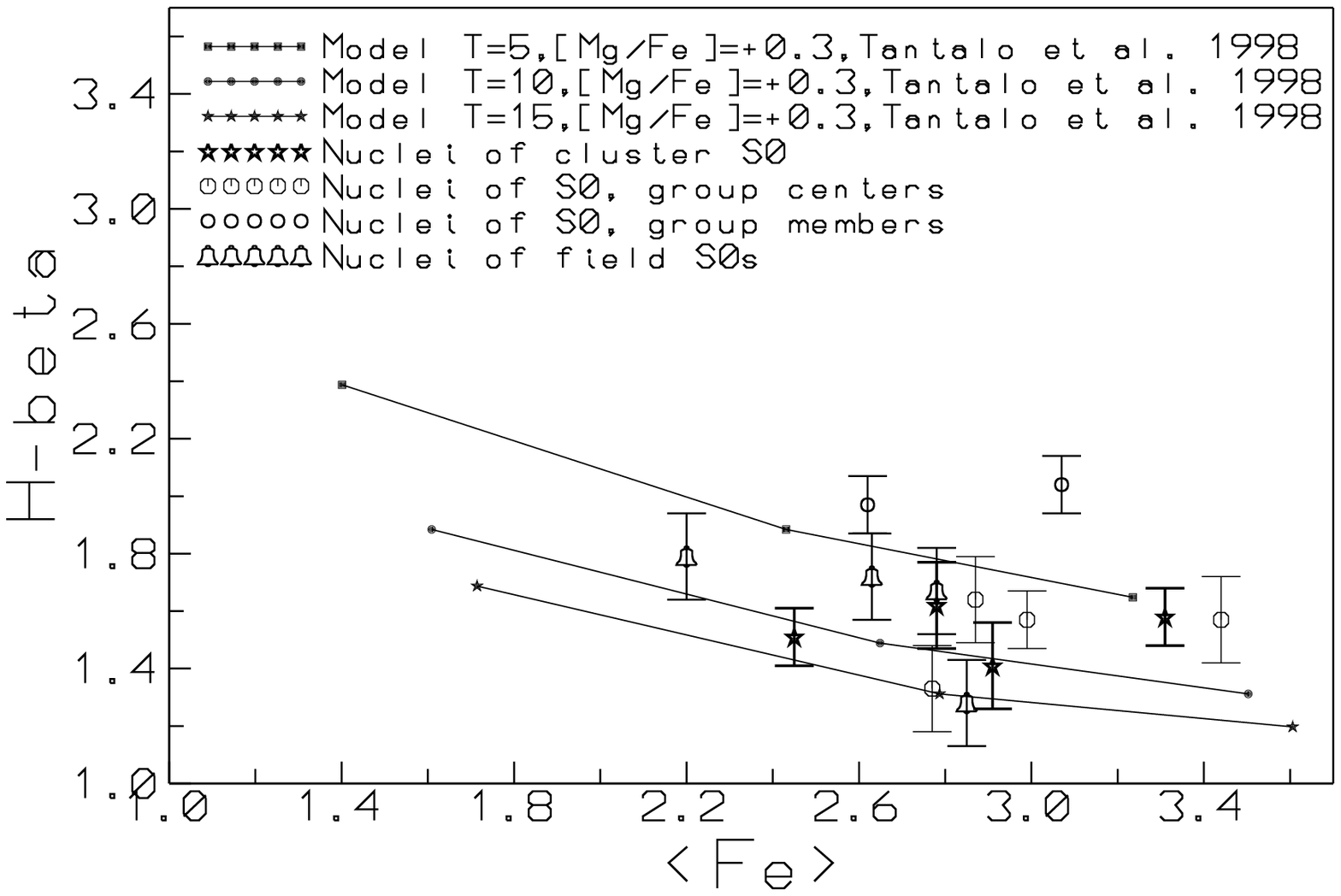}{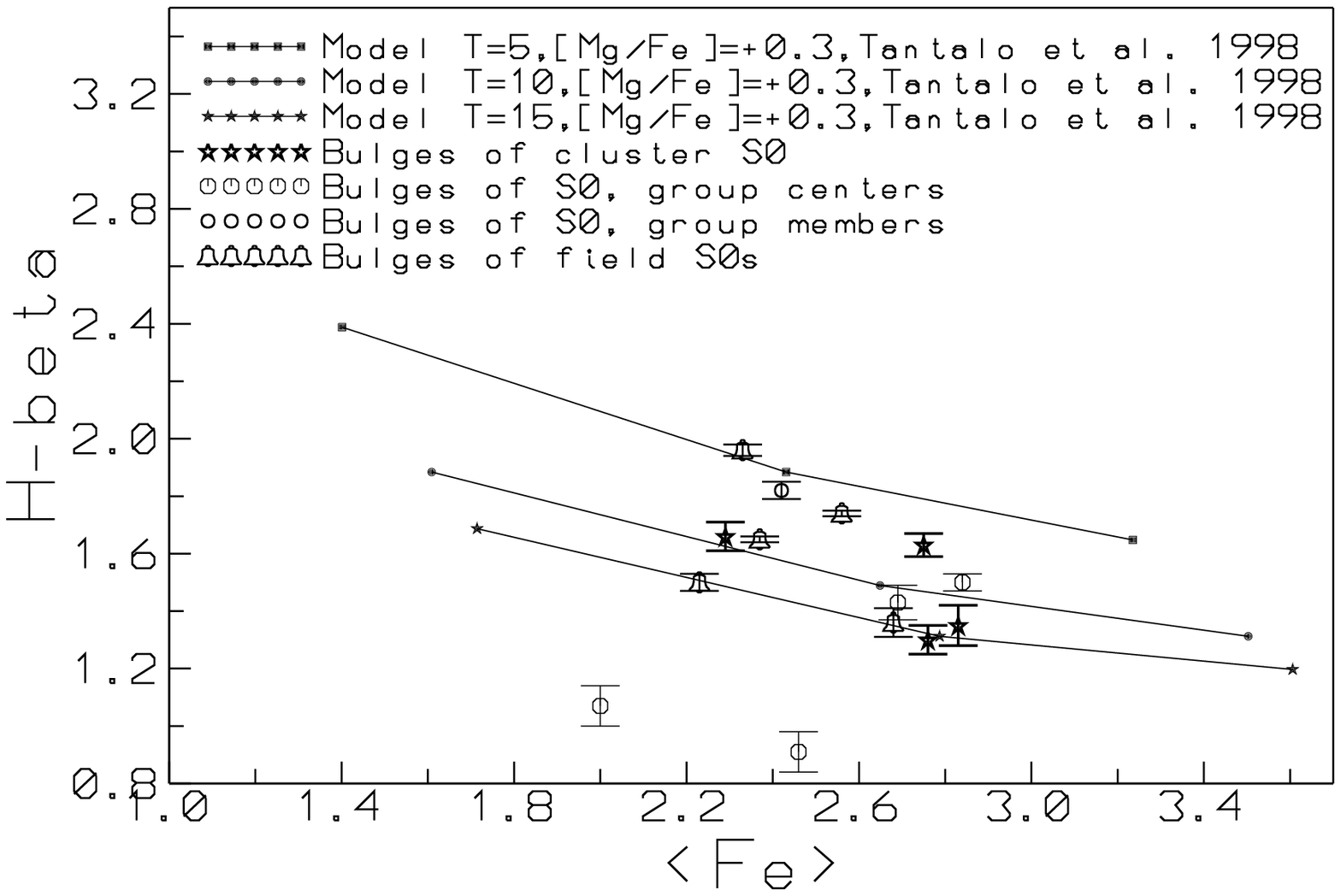}
\caption{The age-diagnostic diagrams for the nuclei (left)
and the bulges (right) of the nearby lenticular galaxies. 
Different
large signs show the galaxies in various environment types. The small
signs connected by thin lines represent stellar population models of equal
ages with [Mg/Fe]$=+0.3$ from Tantalo, Chiosi, \&\ Bressan (1998) (bottom) 
and
 with [Mg/Fe]=0 from Worthey (1994) (top); the signs
-- if one takes them from the right to the left,
mark the metallicities for the Worthey's models +0.50, +0.25, 0.00,
--0.22, --0.50, --1.00,--1.50, --2.00 and for the Tantalo's et al
models +0.4, 0.00, --0.7.}
\end{figure}

As our sample of S0's observed reaches now 35 objects,
we can try to analyze some preliminary statistics concerning mean
metallicities
and stellar ages in the nuclei and in the bulges;
here we would take a `bulge' as a ring with the inner radius of
$3\arcsec$ and the outer radius of $7\arcsec$ so in some cases a
circumnuclear disk may contaminate the bulge measurements. To probe
an environmental effect, we divide our sample into four parts:
cluster galaxies, group central galaxies, peripheric group members,
and field galaxies; in more rough division, the first two categories
can be attributed to the `dense-environment galaxies' and the last
two -- to the `sparse-environment galaxies'. Figure~3 presents
the diagrams confronting two Lick indices, $<\mbox{Fe}>$ {\bf vs}
Mgb, for the nuclei and the bulges separately. One can see
a clear environmental effect for the nuclei: all but one galaxies
in the group centers and all but one cluster galaxies have
[Mg/Fe]$=+0.3 \div +0.4$ in their nuclei, whereas the nuclei of the
field galaxies are uniformly distributed between [Mg/Fe]=0 and
[Mg/Fe]$=+0.3$. Interestingly, for the bulges the environmental
effect is much less pronounced, and in average [Mg/Fe] in the
bulges of the dense-environment galaxies is smaller than in
their nuclei. In general, the bulges of the lenticular galaxies
on the diagram `$<\mbox{Fe}>$ vs. Mgb' are quite different from
the ellipticals of the similar luminosities: they have lower
iron indices and probably lower mean metallicities.

To determine mean (luminosity-weighted) stellar ages, we confront any
metal-line index to the H$\beta$; as we try to take into account also
various magnesium-to-iron ratios in the stellar populations under
consideration, we give two sets of these diagrams, with the models of
Worthey (1994) having [Mg/Fe]=0 (Fig.~4, top) and with models of Tantalo,
Chiosi, \&\ Bressan (1998) for [Mg/Fe]$=+0.3$ (Fig.~4, bottom). From these
diagrams one can see three rather unexpected things: the nuclei of the
lenticular galaxies are obviously younger than their bulges, and some
of them are as young as 1.5--2~Gyr old; though the bulges are in average
older than the nuclei, some of them are as young as 5~Gyr old; and the
magnesium-overabundant nuclei are not so young as the nuclei with
the solar Mg/Fe ratio. The last finding can be easily explained by
the fact known from chemical evolution modeling: a short (shorter
than $10^9$ yrs) star formation burst gives magnesium overabundance,
and to obtain the solar Mg/Fe one needs a continous star formation
during several Gyrs, so in average the solar-Mg/Fe stellar
populations are formed later. Figure~5 gives age
distributions for the nuclear and the bulge stellar populations.
As some faint emission may contaminate H$\beta$ absorption lines
and our age determinations are in fact upper limits, we prefer to
give cumulative distributions, namely, a number of galaxies with
nuclei (bulges) younger than an abscissa age. Again, one can see
the environmental effect for the nuclei: the median age for the
nuclei of the galaxies in sparse environments is 5~Gyr, and the
median age for the nuclei of the galaxies in the dense ones is
7~Gyr, in accordance with the previous claims that the
dense-environment galaxies have higher Mg/Fe in the nuclei and that
the magnesium-overabundant nuclei are in average older. For the
bulges, the environmental effect is absent: the bulges of the galaxies
in any environments have a median age of 10~Gyr.

\acknowledgments
I thank the astronomers of the Special Astrophysical Observatory
V. L. Afanasiev, A. N. Burenkov, V. V. Vlasyuk, S.N. Dodonov, and
the post-graduate student
A. V. Moiseev for supporting the observations at the 6m telescope.
The 6m telescope is operated under the financial support of
Science Ministry of Russia (registration number 01-43).
Also I am very grateful to Dr. R. F. Peletier for the 2D-FIS/WHT
data on NGC 2841. During the data analysis we have
used the Lyon-Meudon Extragalactic Database (LEDA) supplied by the
LEDA team at the CRAL-Observatoire de Lyon (France) and the NASA/IPAC
Extragalactic Database (NED) which is operated by the Jet Propulsion
Laboratory, California Institute of Technology, under contract with
the National Aeronautics and Space Administration.
This research has made use of the ING Archive.
My work is supported by the grant 01-02-16767 of the Russian
Foundation for Basic Researches.

\begin{figure}
\plottwo{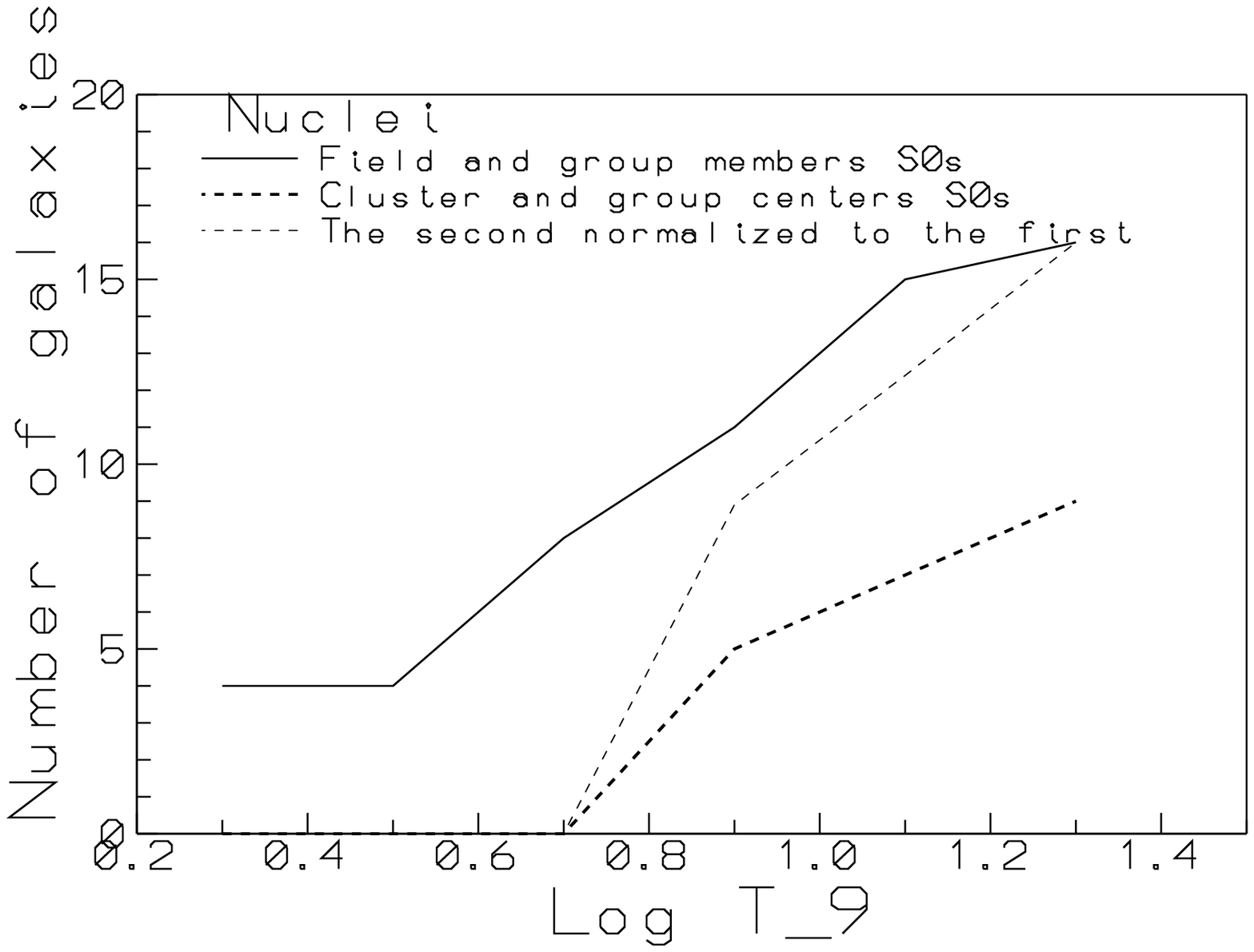}{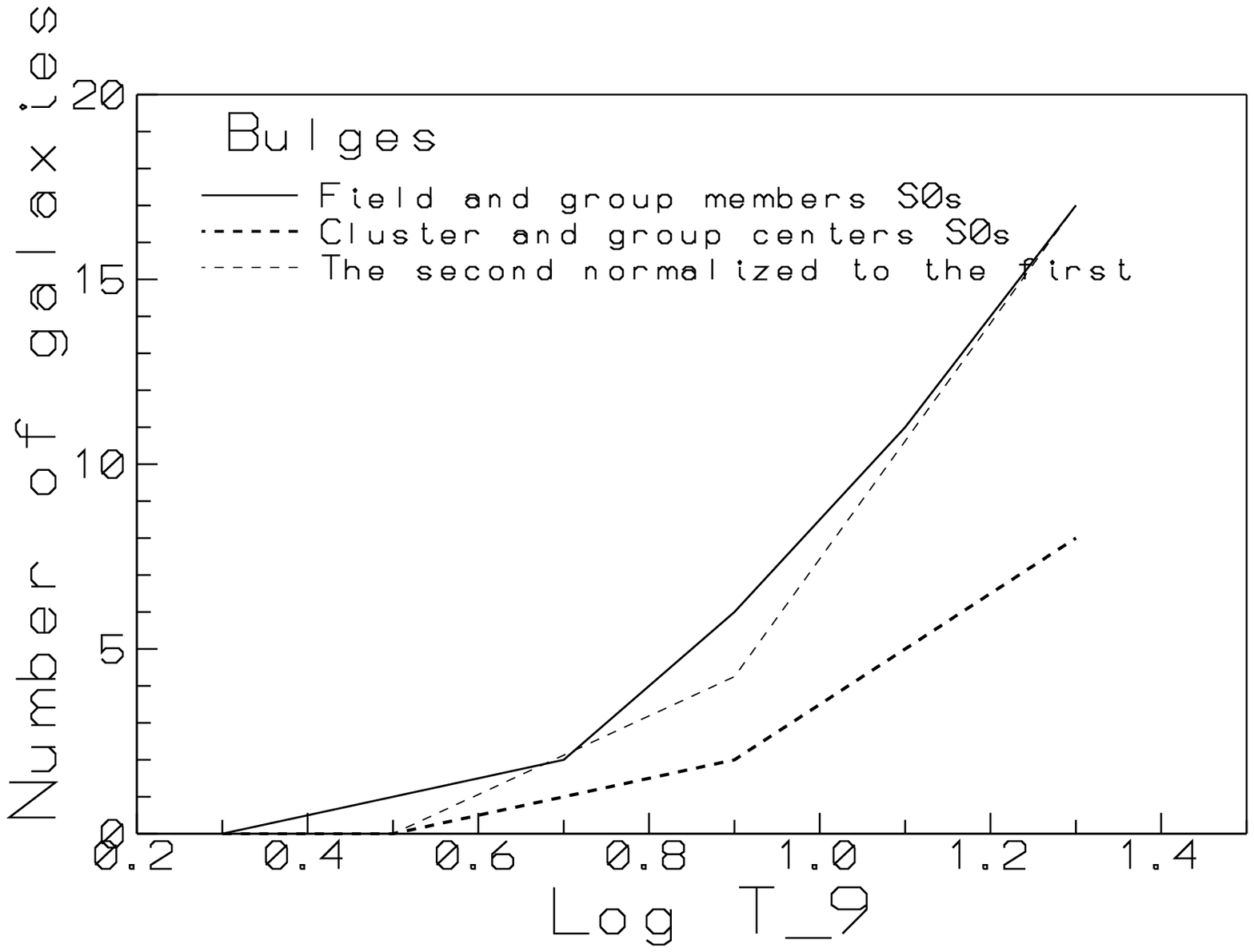}
\caption{Cumulative distributions of the mean stellar ages in
the nuclei (left) and the bulges (right) of the nearby S0
galaxies in sparse environments (solid lines) and in dense
ones (dashed lines).}
\end{figure}


\begin{references}

\reference{Afanasiev, V.L., Sil'chenko, O.K., \&\ Zasov, A.V. 1989,
          \aap, 213, L9}

\reference{Afanasiev, V.L., Vlasyuk, V.V., Dodonov, S.N., \&\ Sil'chenko,
            O.K. 1990, Preprint SAO, N54}

\reference{Afanasiev, V.L., Dodonov, S.N., Drabek, S.V., \&\ Vlasyuk,
        V.V. 1996, MPFS Manual, (Nizhnij Arkhyz: SAO Publ.)}

\reference{Afanasiev, V.L., \&\ Sil'chenko, O.K. 1999, \aj, 117, 1725}

\reference{Bacon, R., Adam, G., Baranne, A., Courtes, G., et
al. 1995, \aaps, 113, 347}

\reference{Bender, R., \&\ Surma, P. 1992, \aap, 258, 250}

\reference{Bertola, F., Buson, L.M., \&\ Zeilinger, W.W. 1992,
               \apj, 401, L79}

\reference{Bertola, F., Cinzano, P., Corsini, E.M.,  Rix, H.-W.,
            \&\ Zeilinger, W.W. 1995, \apj,  448,  L13}

\reference{Fisher, D., Franx, M., \&\ Illingworth, G. 1996, \apj, 459, 110}

\reference{Sil'chenko, O.K., Afanasiev, V.L., \&\ Vlasyuk, V.V. 1992,
      \azh, 69, 1121}

\reference{Sil'chenko, O.K., Vlasyuk, V.V., \&\ Burenkov, A.N. 1997,
       \aap, 326, 941}

\reference{Sil'chenko, O.K. 1999a, \aj, 117, 2725}

\reference{Sil'chenko, O.K. 1999b, \aj, 118, 186}

\reference{Sil'chenko, O.K. 2000, \aj, 120, 741}

\reference{Sil'chenko, O.K., \&\ Afanasiev, V.L. 2000, \aap, 364, 479}

\reference{Sofue, Y., \&\  Wakamatsu, K.-i. 1994, \aj, 107, 1018}

\reference{Tantalo, R., Chiosi, C., \&\ Bressan, A. 1998,
           \aap, 333, 419}

\reference{Vlasyuk, V. V. 1993, Astrofiz. issled. (Izv. SAO RAS) 36, 107}

\reference{Vlasyuk, V.V., \&\ Sil'chenko O.K. 2000, \aap, 354, 28}

\reference{Worthey, G. 1994, \apjs, 95, 107}

\end{references}
\end{document}